\documentclass[12pt,notitlepage]{article}
\usepackage[a4paper,top=25mm,bottom=25mm,outer=25mm,inner=25mm]{geometry}
\usepackage[affil-it]{authblk}

\usepackage{mathptmx}       
\usepackage{helvet}         
\usepackage{courier}        
\usepackage{type1cm}        
\usepackage[format=plain, labelfont={bf}, font={small}, labelsep=quad]{caption}		
\usepackage{float}
\usepackage{setspace}					

\usepackage{amsmath}
\usepackage{makeidx}         
\usepackage{graphicx}        
\usepackage{textcomp, gensymb}        
\usepackage{siunitx}        
\usepackage{multicol}        
\usepackage[bottom]{footmisc}	
\usepackage[colorlinks=true, linkcolor=blue]{hyperref}        
\usepackage{soul}            
\hypersetup{colorlinks=true,urlcolor=blue}

\usepackage{csquotes}
\usepackage[backend=biber,uniquelist=false,maxbibnames=3,giveninits=true,isbn=false,eprint=false,url=false,sorting=none]{biblatex}
\newcommand{\hess}{H.E.S.S.}
\newcommand{\hessi}{H.E.S.S.\ I}
\newcommand{\hessii}{H.E.S.S.\ II}
\newcommand{\nec}{NECTAr}
\newcommand{\lidccd}{LidCCD}

\DeclareSIUnit{\pe}{pe}
\DeclareSIUnit{\cel}{\degree C}

\DeclareUnicodeCharacter{03B3}{$\gamma$}
\DeclareUnicodeCharacter{03B7}{$\eta$}
\DeclareUnicodeCharacter{2212}{--}
\DeclareUnicodeCharacter{0141}{\L}

\makeindex             

\begin{document}
\title{H.E.S.S.: The High Energy Stereoscopic System}
\author{Gerd Pühlhofer\thanks{corresponding author; email: \href{mailto:gerd.puehlhofer@astro.uni-tuebingen.de}{gerd.puehlhofer@astro.uni-tuebingen.de}} ~and Fabian Leuschner and Heiko Salzmann}
\affil{Institute for Astronomy and Astrophysics T{\"u}bingen\\T{\"u}bingen, Germany}

%
%
\maketitle

\begin{abstract}
	The High Energy Stereoscopic System \hess\ is an array of Cherenkov Telescopes located in the Khomas Highlands in Namibia. \hess\ started operations in 2003 and has been operated very successfully since then. With its location in the Southern hemisphere, the system provides a privileged view of the Milky Way and the Galactic center region. With \hess, a large variety of new TeV emitters has been discovered, both in our Galaxy and in extragalactic space. We provide a description of the individual telescopes and of the system as a whole, and review the scientific highlights that have been achieved with the instrument.
\end{abstract}


\setcounter{tocdepth}{2}
\tableofcontents

\section{Introduction}\label{intro}

Few years after the first detection of a non-thermal $\gamma$-ray source with an Imaging Atmospheric Cherenkov telescope (IACT) at the Mt. Whipple observatory, the first detectors operated by European institutions detected very-high energy (VHE) photon sources as well. 
Most noticeable were the High Energy Gamma Ray Astronomy (HEGRA) array \cite{hegra-crab-1996} and the Cherenkov Array at Themis (CAT) imager \cite{cat-system-1998}. 
The former was an array of IACTs located at the Observatorio del Roque de los Muchachos on the Canary island La Palma, the latter a single IACT in the French Pyrenees. 
With HEGRA, for the first time ever stereoscopic IACT observations -- that is, with at least two telescopes at the same time -- were successfully conducted and proven to improve air shower reconstruction \cite{hegra-stereo-1996}.\par
Already in 1997, the idea emerged among HEGRA scientists to build an array of larger IACTs. 
This plan was logged in a letter of intent \cite{hess-loi}.
The telescopes should be able to detect fainter sources than at that moment possible, such as Galactic Supernova remnants or extragalactic sources out to a redshift of $z\approx 1$.
Following the letter of intent, the initial \hess\ collaboration formed, consisting of scientists and engineers from Germany, France, the UK, Ireland, the Czech Republic, Armenia, South Africa and Namibia.
In summer 2002 the first of the initial four telescopes was operational. 
The remaining three followed until December 2003, forming the \hess{} phase I (\hessi) array.
They were designed to have an energy threshold of $\SI{100}{\giga\eV}$.\par
The stereoscopic approach developed within HEGRA is a key feature of the \hessi{} array:
If a Cherenkov shower is detected with one telescope only, the accuracy to reconstruct its 3D-geometry and therefore the origin and energy of the initial particle has limitations.
However, if two or more telescopes detect the shower, it can be reconstructed in 3D, allowing to obtain the above-mentioned quantities much more precisely \cite{hegra-stereo-1996}.\par
\begin{figure}[ht]
    \centering
    \includegraphics[width=\textwidth]{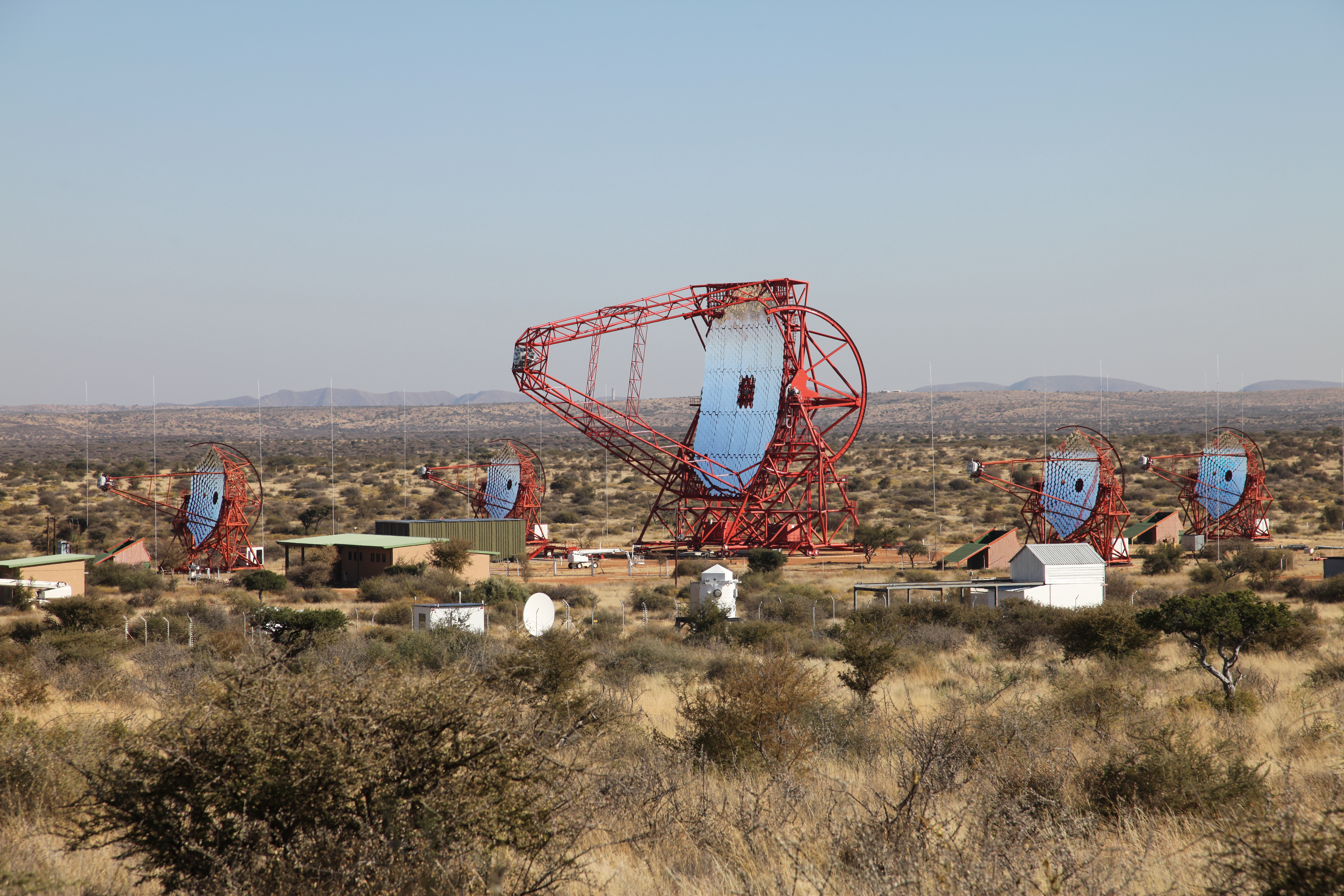}
    \caption{The \hess\ array with the four \SI{12}{m} telescopes CT1-4 and the $\SI{28}{m}$ telescope CT5 in the center. (Reprinted with kind permission of the \hess\ collaboration/Frikkie van Greunen)}
    \label{fig:frikkie_array}
\end{figure}

 
Less than two years after completion of the initial four telescopes (CT1-4), which have mirror dish diameters of \SI{12}{m}, the original plan to build another twelve of these telescopes was dropped.
Instead, the decision was made to design and construct a single, very large telescope in the center of the array, the CT5 telescope.
With a mirror area of \SI{600}{\square\m}, it allows observations in an energy range down to \SI{20}{\giga\eV}. 
This improves the overlap between IACTs and space-born missions, mainly the Fermi Gamma-Ray Space Telescope (at that time known as ``GLAST detector'') reaching up to at least \SI{100}{\giga\eV} \cite{expansion-punch-2005}.
The resulting array of five telescopes is referred to as \hessii{} and is depicted in \autoref{fig:frikkie_array}.
Due to the lower energy threshold and the advanced camera, the event rate increased by more than an order of magnitude compared to the initial array. 
The threshold permitted, e.g. to detect pulsations from the high end of GeV-detected pulsar spectra.
The overall science case justified to accept the worse angular resolution of the (single) large telescope, compared to the \hessi{} array.\par
Today, the \hess\ collaboration consists of more than 200 scientists and engineers in 13 countries. 
At the time of writing, it operates the only IACTs in the southern hemisphere.
Furthermore, it is the only IACT array operating telescopes of different sizes and design simultaneously.
In 2009, a study ranked \hess\ among the ten most influential observatories in the world \cite{telescope-influence-2009}.
After the initially planned end of operations in 2019, \hess\ is currently in its second extension phase, lasting until 2024.
As predictably no facility south of the equator will be nearly as sensitive to $\gamma$-rays by then, chances are good to continue operations after 2024 as well.

\section{The H.E.S.S. Telescopes in Namibia}

\subsection{H.E.S.S. Site}

Several factors play a key role in the decision for a telescope site to perform astronomical observations. Firstly, the site should offer little to no light pollution from nearby civilization. Secondly, a (comparably) good connection to infrastructure (e.g. roads) is helpful for the transport of technical equipment to the site. Thirdly, and most important, a good climate providing low cloud coverage and little rain over the course of a year is a mandatory requirement in order to lose as little observation time as possible due to environmental conditions. Furthermore, the optimal altitude for an array of IACTs is about $\SI{1500}{m}$ resulting from a trade-off between the atmospheric transparency to Cherenkov light and the shower development through the atmosphere \cite{deNaurois2015}. An additional consideration concerning the science cases was that the southern hemisphere provides an excellent view in the central part of our Galaxy. Last but not least, the chosen country should be politically stable to operate the array in a safe environment. \par 
The decision for the \hess\ site fell towards the Khomas highland in Namibia, which is renowned for its good observational conditions and was also considered as a possible site for the European Southern Observatory (ESO) optical telescopes. It offers a mild climate with temperatures ranging from \SI{0}{\cel} to \SI{35}{\cel}, little rain and no snow allowing the telescopes to be operated without protective enclosures. Furthermore, as it is $\SI{100}{km}$ southeast of the capital city of Windhoek it provides good connectivity to the next-closest international airport. The exact location of the telescopes is at $23^{\circ}16'18''$ S and $16^{\circ}30'00''$ E on the Farm G{\"o}llschau at an altitude of about $\SI{1800}{m}$ \cite{hofmann2001}. \par 
The telescope site offers a control building and multiple halls for the storage of spare parts and technical equipment. The control building includes the control room from which the telescopes are operated. Furthermore, it contains the computer cluster handling the science data, a handful of offices and a mechanical workshop for small repairs. A few hundred meters away from the telescope site the residence building is located. It offers accommodations for the shift crew conducting night shifts as well as for the permanent technical staff.

\subsection{Telescope Optical Systems}

The task of the optical system of an IACT is to collect the faint Cherenkov light generated by electromagnetic air showers and reflect it into the UV-sensitive cameras installed in the telescope's focal plane. Most of the currently operating IACT experiments accomplish this with single reflectors, which have a tessellated mirror design in order to have collection areas of $\geq \SI{100}{m^2}$. A tessellated mirror design reduces the cost as well as the optical quality requirements on the single mirror facets compared to a single, large-sized mirror. \par

This section covers the various aspects of the optical systems of the five \hess\ telescopes. 

\subsubsection{Telescope Structures and Drive Systems}

The structures of all five \hess\ telescopes have an altitude-azimuth mount and are made out of steel. The elevation axis ends in two towers, which can be moved in azimuth on a rail with a diameter of $\SI{13.6}{m}$ and $\SI{36}{m}$ for CT1-4 and CT5, respectively \cite{bernloehr2003, hofverberg2013}. The steel structures provide the necessary stiffness to minimize deformations of the reflector shape over a large range of elevation angles. In particular, with this an elevation-dependent adjustment of the mirror alignment is not necessary. \par 
The $\sim \SI{60}{t}$ telescope structure of each of the CT1-4 telescopes is moved on six wheels with friction drives in azimuth with a range of $\geq 385^{\circ}$. The same drive system is used for the elevation drive, which offers a range from $-35^{\circ}$ to $90^{\circ}$ with respect to the horizontal position. In both axes, the telescopes slew at a velocity of $\SI{100}{^{\circ}\per\min}$ \cite{bolz2004}. The $\sim \SI{600}{t}$ structure of CT5 is moved in azimuth on twelve wheels with a total motion range of $560^{\circ}$. The twelve wheels are powered by four servo motors. The same amount of servo motors are used to move the telescope around the elevation axis within a motion range from $-32^{\circ}$ to $175^{\circ}$ with respect to the horizontal position. The velocity of the azimuth and elevation axis is $\SI{200}{^{\circ}\per\min}$ and $\SI{100}{^{\circ}\per\min}$, respectively \cite{deil2008, hofverberg2013}.

\subsubsection{Mirror Systems}

There are two basic approaches in the design of a single IACT reflector geometry. In the Davies-Cotton design, each spherical mirror facet has the same focal length $f$ and the mirrors are arranged on the mirror support structure according to a sphere with a radius of \textit{f}. The mirror facets' optical axes are aligned to a distance of $2f$. The advantage of the Davies-Cotton design is that it provides a sufficiently good optical point spread function (PSF) even for light rays which have a large incident angle to the optical axis (off-axis light rays). The disadvantage is that it is not isochronous resulting in a spread of photon arrival times in the telescope's focal plane \cite{aharonian2008}. This is not the case for a parabolic layout in which the (usually spherical) mirrors are arranged on a paraboloid with $z= r^2/(4f)$. In an ideal parabolic layout, the focal length of each mirror facet varies depending on its distance from the center of the reflector. The better isochronicity comes at the expense of a worse off-axis PSF compared to the Davies-Cotton design \cite{bernloehr2003}.  \par 
The tessellated mirrors of the CT1-4 telescopes consist of 380 round mirror facets with a diameter of $\SI{60}{cm}$ each, resulting in a total mirror area of $\SI{107}{m^2}$. The reflectors follow a Davies-Cotton layout with a diameter of $\sim \SI{12}{m}$ and a focal length of $\SI{15}{m}$ leading to a $f/d \approx 1.2$. For a reflector with this focal length and diameter this results in a time dispersion of $\sim \SI{1.4}{ns}$. This is still uncritical as it is in the same order of magnitude as the intrinsic time spread of the air shower photons \cite{bernloehr2003}. \par 
On the other side, the reflector of the large telescope (CT5) follows a parabolic layout with a rectangular shape of $\SI{24}{m}$ x $\SI{32}{m}$ (width x height), equivalent to a circular reflector with a diameter of $\SI{28}{m}$. The tessellated mirror consists of 876 hexagonal mirror facets with a flat-to-flat diameter of $\SI{90}{cm}$, resulting in a total mirror area of $\SI{614}{m^2}$ \cite{gottschall2015, cornils2005}. The focal length of the reflector is $\SI{36}{m}$ with a $f/d$-ratio of $\sim 1.3$ \cite{hofverberg2013}. A parabolic layout was chosen because for a Davies-Cotton layout the time dispersion grows linearly with the reflector diameter for a fixed $f/d$-ratio. In the case of a reflector diameter of $\SI{28}{m}$, this would have led to an increased energy threshold as only more energetic $\gamma$-ray showers would have triggered the camera. As mentioned above, in an ideal parabolic layout the focal length of the mirror facets varies with the distance to the center of the reflector. However, it was shown with simulations that choosing mirror facets with the same focal length has a negligible impact on the imaging qualities of the telescope and thus does not warrant the higher manufacturing costs \cite{cornils2005}, and therefore all mirror facets have equal focal length. \par

\subsubsection{Mirror Alignment}

The objective of the mirror alignment is to focus the light reflected by each mirror facet into a single focus spot. Therefore, the mirror alignment defines the position where the telescope's optical axis crosses the telescope's focal plane and it is closely linked to the optical psf and the pointing accuracy of the telescopes. The optical psf quantifies how good a point source is depicted in the telescope's focal plane, whereas the pointing accuracy describes how well the telescope's optical axis aligns with the direction to the observed source. A proper alignment of all mirror facets is therefore essential for a good quality of the scientific data. Both of these quantities, the optical psf and the pointing accuracy, will be explained in more detail in \nameref{psf-pa}. This paragraph focuses on the mirror alignment control systems and the mirror alignment procedure itself. They are described following \cite{bernloehr2003, cornils2003} and \cite{gottschall2015} for CT1-4 and CT5, respectively.\par

\begin{figure}[ht]
    \centering
    \includegraphics[width=.7\textwidth]{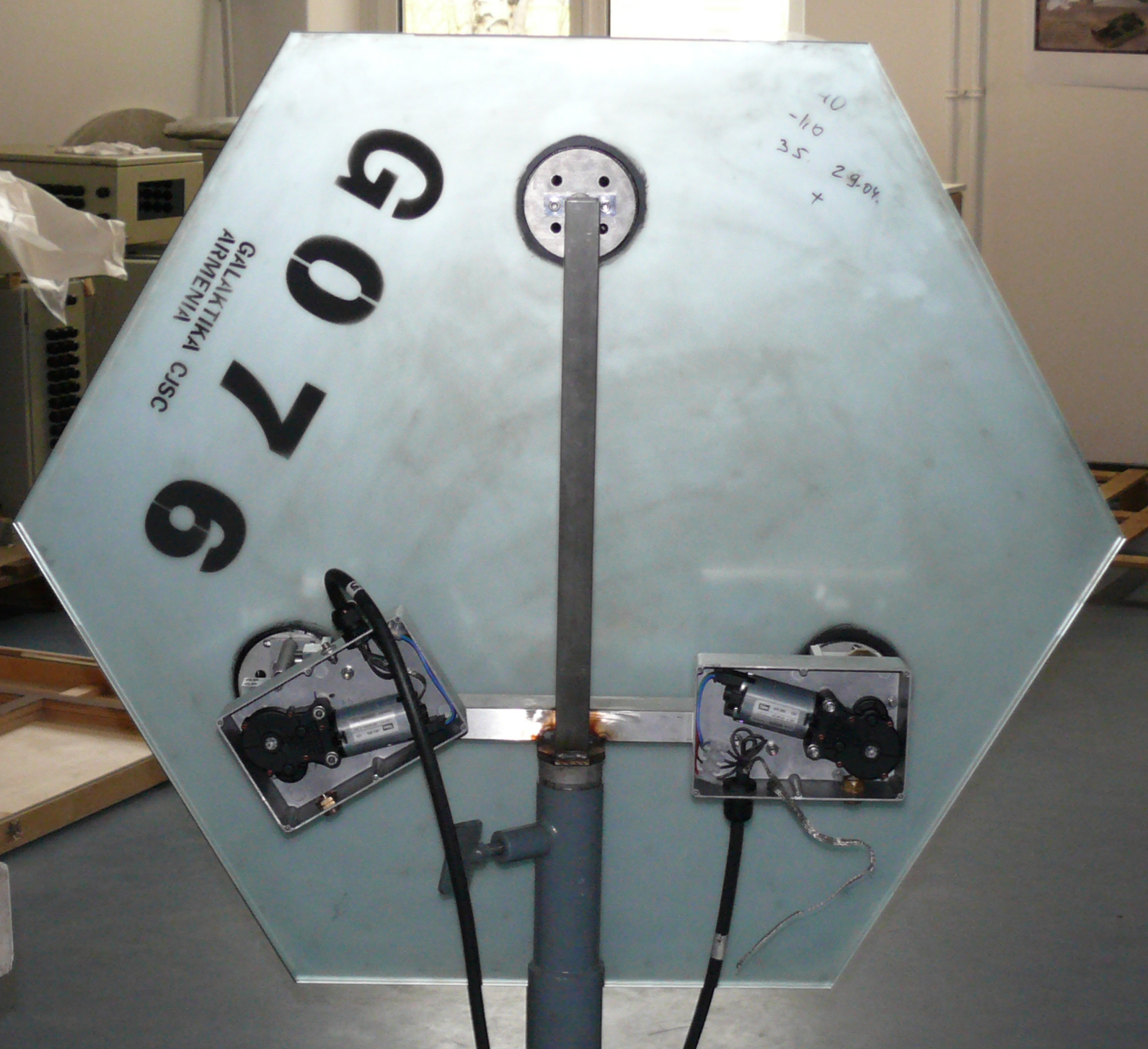}
    \caption{The backside of a mirror for CT5. The fix point is located at the top, whereas the actuators are located at the lower left and lower right part of the mirror. The motor boxes of the actuators are opened and reveal a look at the motors that drive the actuators. (Reprinted with kind permission from IAAT/Thomas Schanz)}
    \label{fig:mirror_backside}
\end{figure}

While the technical details between the mirror alignment system for CT1-4 and CT5 differ, the basic approach and working principle is the same. As is depicted in \autoref{fig:mirror_backside}, each mirror facet is mounted to the mirror support structure via two motorized actuators and a fix point. They are screwed to metal plates glued to the back of the mirror. In contrast to the fix point, the actuators are adjustable in length and are used to regulate the tilt of the mirror. 
The actuators of CT1-4 provide a minimal step size of $\SI{3.4}{\mu m}$, which yields an accuracy of the mirror tilt of $\SI{0.013}{mrad}$ ($\sim \frac{1}{250}$ of a camera pixel FoV). The actuators of CT5 have a minimal step size of $\SI{2.4}{\mu m}$, leading to a mirror tilting accuracy of $\SI{0.011}{mrad}$ ($\sim \frac{1}{100}$ of a camera pixel FoV). \par
Mirrors are aligned such that they provide the best imaging quality in the main observation range of the telescopes. For this, a star is tracked between an elevation angle of $60^{\circ}$ and $70^{\circ}$. The star light of each mirror facet is reflected onto an optical target (i.e. a diffusively reflecting surface), which is placed in front of the camera. For the CT1-4 cameras the optical target is the camera lid and for the current CT5 camera (FlashCam) it is a dedicated circular plate with a diameter of $\SI{175}{mm}$. The necessary optical feedback about the star spots reflected by the mirror facets is provided by a charged-coupled device (CCD) camera installed in the middle of the mirror support structure. This CCD camera is dubbed as \lidccd{} as it observes the camera's lid. If the mirrors are unaligned, each mirror facet causes a single star spot in the \lidccd{} image. In order to derive the correlation between the movements of the actuators for a single mirror and the resulting movement of the reflected star spot, the \lidccd{} takes an image before and after each of the actuators is moved by a certain amount of steps. From the difference of the before and after images, the movement of the star spot can be determined and a correlation between actuator and star spot movement can be derived. Using this calibration, the mirror can be moved to the required position. This procedure is repeated for each of the mirror facets until the light of all mirror facets is reflected into a single spot. The mirror alignment itself is mostly automatized.\par 
In case the telescope structure would deform too much under different elevation angles due to the gravitational load and therefore the imaging qualities would deteriorate, an active alignment, i.e. an alignment of the mirrors depending on the elevation angle, would be necessary. However, the steel structure of the \hess\ telescopes is stiff enough to keep the imaging quality over the whole observational elevation range at an acceptable level and therefore a single initial alignment setting is used for the whole elevation angle range. 

\subsubsection{Point Spread Function and Pointing Accuracy}\label{psf-pa}

The optical PSF and the pointing accuracy are two of the main characteristics of a telescope's optical system. Both of these quantities have an impact on the scientific data taken with the telescope. While the optical PSF describes how good a point source is depicted in the telescope's focal plane and therefore quantifies the air shower imaging capabilities of the telescope, the pointing accuracy describes how accurate the telescope's optical axis aligns with the direction to the observed astrophysical source. The latter is critical in determining the direction of a Cherenkov shower and therefore the position of the astrophysical source. Both, optical PSF and pointing, need to be good and well-understood in order to ensure the best scientific data quality. \par 

The shape and size of the optical PSF depends on various parameters. Among them are the reflector geometry (impacting e.g. the off-axis PSF), the optical quality of the single mirror facets (e.g. the reflectivity, size of the mirror facet PSF) as well as the quality of the mirror alignment (e.g. the positioning uncertainty of the actuators and therefore the mirror tilts). In \hess, the on-axis PSF is monitored regularly. The off-axis psf is linked to the on-axis psf via the reflector geometry and therefore the monitoring of the on-axis psf is sufficient to quantify possible deterioration of the mirror alignment or the telescope structure. For past measurements of the off-axis PSF for CT1-4 and for simulations of the off-axis PSF of CT5 the reader is referred to \cite{cornils2003} and \cite{cornils2005}, respectively. \par
The procedure to measure the PSF is the same for all of the five telescopes. The PSF is measured as a function of the elevation angle to quantify the deformations of the mirror support structure under different gravitational loads. This is accomplished by tracking $~30-40$ stars distributed between an elevation angle of $20^{\circ}$ and $90^{\circ}$, while the \lidccd{} takes images of the light reflected on the optical target of the camera. Starting from the center of gravity of the star spot in the \lidccd{} image, the radius $r_{80\%}$ containing $80\%$ of the light intensity is calculated and serves as a figure of merit for the PSF. The distribution of $r_{80\%}$ as a function of the elevation angle is fitted with a model describing the bending of the mirror support structure. Furthermore, the PSF at the designed optimal elevation angle of $65^{\circ}$ is extracted from the fitted model to monitor the long-term temporal evolution of the PSF on a per telescope basis. For all of the five telescopes, the optical PSF is well below the respective camera pixel size and the PSF proved to be sufficiently stable as a function of time \cite{cornils2005, gottschall2015}. Therefore, no realignment of mirrors was performed over the lifetime of \hess, except in the case of mirror facet exchanges. Furthermore, no mirror alignments are planned in the remaining lifetime of \hess, and the observed small and steady deterioration of the PSF will have no substantial impact on the scientific performance of the instrument. \par 
For the pointing, the main goal is to understand the behavior of the telescope's optical axis as a function of the telescope position in elevation and azimuth. Ultimately, the achieved pointing accuracy determines how well a $\gamma$-ray point source with high enough photon statistics can be localized in the sky. While single $\gamma$-ray showers are reconstructed with typical accuracies of $0.1^{\circ}$, source localization of tens of arcseconds can be achieved. (e.g. \cite{puehlhofer1997}). \par 
The pointing accuracy is determined by the tracking accuracy of the telescopes, as well as by the bending of the mirror support structure (including the actuators). The tracking accuracy of CT1-4 has a root mean squared in the order of $ 0.9''$ in elevation and $1.9''$ in azimuth, while it is for CT5 slightly larger with $2.1''$ in elevation and $3.0''$ in azimuth \cite{bolz2004, hofverberg2013}. This is below the achievable pointing accuracy and can normally be ignored. Bending effects can be calibrated offline, since most effects are elastic and therefore reproducible. The bending (including also the residual misalignment of the focus spot) is regularly corrected in data analysis with a bending model of the telescopes, the parameters of which are periodically (every few months) determined with regular calibration runs, employing the same machinery as for the PSF monitoring. With these pointing models, pointing accuracies in the order of  $\leq 20''$ for CT1-4 and $\leq 60''$ for CT5 can be achieved.\par 
For special applications, a mode called precision pointing has also been employed, in which the bending model was partially replaced by information from CCDs located in the telescopes' dishes that monitor the optical sky around the Cherenkov cameras \cite{precisionpointing2010}. Here, even higher pointing accuracies have been achieved.

\subsection{Cameras}\label{secCameras}
The cameras of the \hess\ array must record the short and faint Cherenkov light flashes induced by air showers.
While CT1-4 share the same camera, CT5 has a different one installed. 
After the inclusion of CT5 into the system, in order to permit an effective hybrid trigger (see \nameref{CentFac}), the readout dead-time of the original CT1-4 cameras became a bottleneck.
This was remedied with the upgrade of the CT1-4 camera electronics system, which employed the New Electronics for the Cherenkov Telescope Array (\nec) technology that was originally developed for CTA \cite{NECTAr}.
In 2015, the first of the upgraded cameras was installed, followed by an intensive commissioning phase. 
In September/October 2016, the other three cameras were upgraded and started science operations soon after. 
One refers to these cameras as \textit{HESSIU}, short for ``H.E.S.S.-I-Upgrade''.\par
The need for faster electronics wasn't the only reason to update the cameras. 
Multiple years out in the Namibian savanna left their mark and while the risk of failures increased, maintenance became more and more difficult as electronic components became obsolete. 
The same happened to the initial camera of CT5, which suffered the conditions even more as there is no shelter for the big telescope's camera. 
In October 2019, the CT5 camera was exchanged with the first fully functional FlashCam Camera, which is based on another camera technology developed for CTA \cite{FlashCam2019}.
This section gives an overview about the technical aspects of the cameras used on site as of writing of this document (2023).

\subsubsection{CT1-4: The HESS1U Cameras}

\begin{figure}
    \centering
    \includegraphics[width=\textwidth]{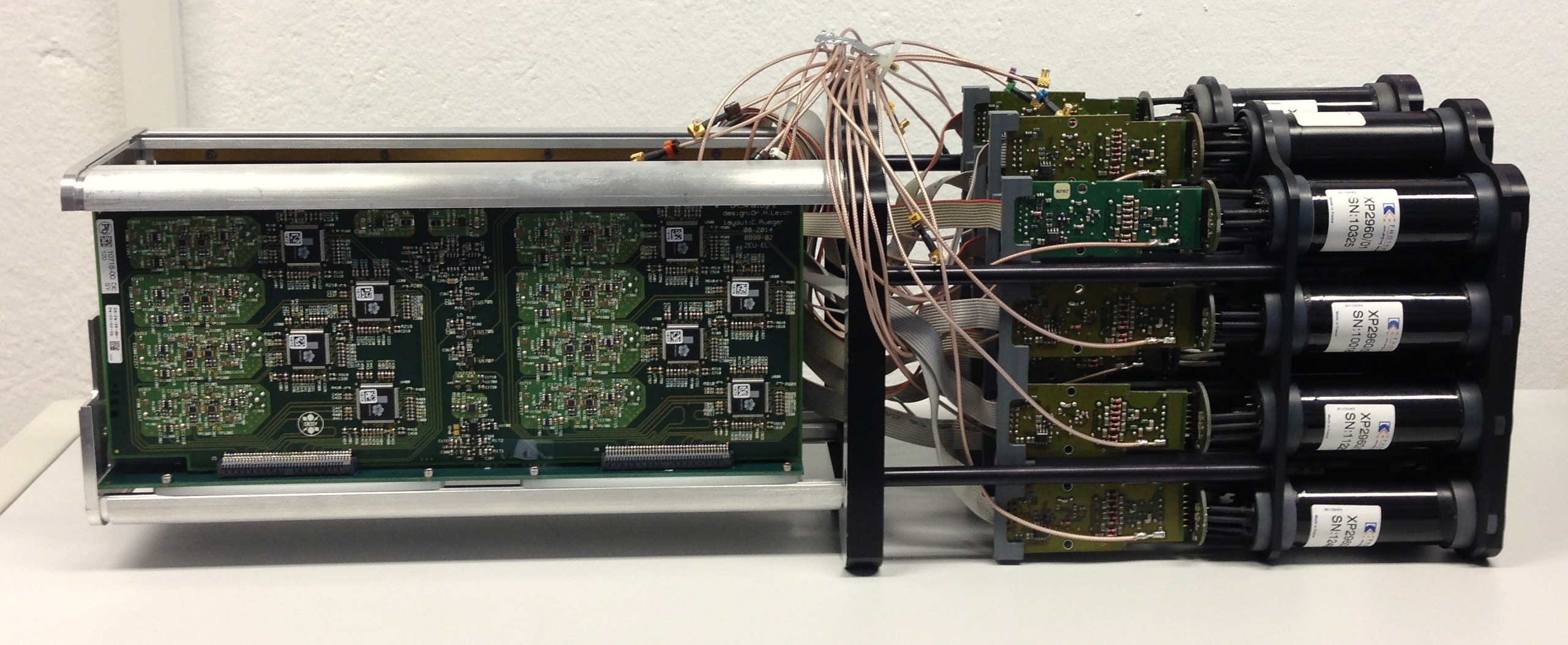}
    \caption{A single drawer of the HESS1U cameras. On the right, the first row of the 4x4 pixel matrix is visible including HV bases. The left (rear) half of the drawer consists of the slow control board (bottom, not visible) and the two analogue boards connected orthogonally to it. One of the analogue boards with eight \nec\ chips (black, squared chips with QR-code) can be seen. On top, the signal cables are bundled together. In assembled state, they are connected to the analogue boards on the inside. Additionally, when assembled, a Faraday cage protects the drawer electronics against electronic noise. (Reprinted with kind permission from DESY / Stefan Klepser)}
    \label{fig:hess1u-drawer}
\end{figure}

A detailed description of the HESS1U cameras is given in \cite{HESS1U}. 
This paragraph is mainly based on the information provided by that source.\par
Each of the four HESS1U cameras consists of 960 pixels.
As light detectors, photomultiplier-tubes (PMTs) are used.
Winston cones on top of the PMTs minimize both, light loss due to pixel spacing and contamination by background light from large straying angles.
The PMTs together with the cones have an FoV of \ang{0.16} each.
While the camera electronics was completely renewed during the upgrade, the PMTs and their high voltage power supplies (\textit{HV bases}) are still the original ones, as they are robust on the one hand and expensive on the other hand.\par
Key building blocks of the HESS1U cameras are the drawers.
There are sixty of them, each holding sixteen PMTs at the front end.
\autoref{fig:hess1u-drawer} shows a picture of one drawer.
The HV bases are connected to a slow-control board, which is equipped with a field programmable gate array (FPGA) and an advanced Reduced-Instruction-Set-Computer machine (ARM) single board computer and which is in charge of controlling the whole drawer.
For handling the PMT output, each drawer is equipped with two analogue boards, each holding eight channels, one for each PMT in the drawer.
A support structure mechanically stabilizes the drawers.
It fastens the PMTs with their HV bases at the front end, then the slow-control and analogue boards and is completed by a connection board that holds sockets for communication between the drawer and the rest of the camera, as well as for the power supply of the whole drawer.\par
The output of each PMT is sent to the respective channel on the analogue board, where the analogue signal processing takes place.
The signals are first pre-amplified and then split into three branches.
One of these branches directs to a high-speed comparator, whose digital output is used by the FPGA on the slow control board for the drawer trigger decision. 
This is called the level 0 (L0) trigger signal and is further processed to make the camera trigger decision.\par
The other two branches are routed to the two inputs of the \nec\ chip.
One of them is the high gain, the other one the low gain channel for weak and strong signals, respectively.
After amplification, the signals in both readout channels are stored individually in two 1024 cell analogue ring memory buffers with a switching frequency of \SI{1}{\giga\Hz}, i.e., each recorded charge is stored for \SI{1024}{\nano\s} before being overwritten.\par
Upon a camera-wide trigger signal, the charges in a defined region of interest (typically sixteen cells, i.e. \SI{16}{\nano\s}), are read out.
They are digitized with an on-chip analogue-to-digital-converter (ADC) and sent to the FPGA on the slow control board.
The FPGA normally sums the sixteen charges up to the integrated charge value of the respective event\footnote{\textit{Event} is the general term to refer to a Cherenkov shower that is recorded by any of the cameras.} and pixel.
This value is buffered in the ARM computer and eventually sent to the camera server, which is located at the ``computing farm'' in the control building (see \nameref{CentFac}).\par
All drawers are mounted to one metal plate which separates them from the camera back-end electronics of which the drawer interface box (DIB) is the most important part.
It is basically the ``brain'' of the camera, holds and distributes a central clock to all drawers, processes the various L0 trigger towards a camera trigger, controls the readout and provides an interface to the entire array control.\par
The clock of the camera is a \SI{10}{\mega\Hz} quartz oscillator, synchronized to a GPS-based pulse-per-second signal.
It is used to timestamp events at the camera level with a precision of a few \si{\micro\s}, which is about an order of magnitude faster than the typical trigger rate of the cameras.
Since early 2022, the HESS1U cameras are additionally connected to the array-wide White Rabbit (WR) timing system that was implemented together with the FlashCam camera at CT5.\par
Camera trigger decisions are made in the DIB, aided by a fast-sampling FPGA.
For the trigger decision, the camera is divided into 38 sectors of 64 pixels each, horizontally overlapping by a half and vertically overlapping by a complete drawer.
In short, the standard trigger logic works as follows: the camera is triggered if $N$ pixels within one sector are brighter than a defined threshold $P$, given in photo-electrons (\si{\pe}).
Usually, $N=3$ and $P=\SI{5.5}{\pe}$.
Once the camera has triggered, all the drawers are read out and a signal is sent to the data acquisition system (see \nameref{CentFac}) to indicate the trigger. 
As this logic is implemented exclusively on an FPGA, it could be changed to other solutions, if this was desired.\par
Other important parts of the back-end electronics are the ventilation and heating system, the power distributors and the pneumatic system, which is responsible for opening and closing the camera lid and the rear door.
The ventilation and heating system consists of a single, \SI{250}{\milli\m} fan and a \SI{6}{\kilo\W} heater.
The fan produces an airflow of $\sim$\SI{360}{l/s} from the back to the front of the camera.
Filters ensure that as little dust as possible enters the camera body.
The heater starts automatically when the external, relative humidity exceeds 75\,\% or when the outside temperature drops below \SI{5}{\cel}.
This prevents condensation inside the camera and minimizes the temperature gradient across the camera, respectively.
With this system, a stable temperature of \SI{32}{\cel} with a gradient of $\pm$\SI{5}{\cel} is achieved inside the camera during operations. 
There is no measurable effect on the data and trigger efficiency due to the temperature and its gradient.\par
The power system is mainly based on off the shelf solutions.
Only the \SI{24}{\V} DC required by the drawers is distributed from the main power supply via a custom-built solution, the Power Distribution Box (PDB).
Besides supplying the drawers with power, it also monitors their consumption and shuts them off autonomously in case an over-current occurs.
The entire camera's power consumption is about \SI{3}{\kilo\W} in basic operation, plus possibly \SI{6}{\kilo\W} if the air heater is turned on.\par
To handle computations on the \si{\mu\s} level, all camera subsystems are connected via Ethernet switches inside the camera.
Access from outside is possible via the camera server, which is linked to the Ethernet switches as well using a \SI{10}{Gbit/s} optical fiber.
It is this server that issues all necessary slow-control commands.
However, during data-taking the camera works quite independent, i.e. all data is buffered in the DIB.\par
In conclusion, the deployment of the HESS1U cameras in 2015 and 2016 provided a significant improvement for operations of the H.E.S.S. experiment.
Due to the reduced dead time by a factor of $\sim60$ compared to the original CT1-4 cameras, the number of events recorded hybrid (that is, at least one telescope of CT1-4 together with CT5) almost doubled.
From the upgrade until 2020, a data taking efficiency of 98.5\,\% was obtained on average for CT1-4. 
Also after that, when the Covid-19 pandemic started and European maintenance teams were not allowed to enter Namibia, the HESS1U-cameras kept working with great success. 
The cameras are expected to work successfully until the end of the lifetime of the \hess experiment.

\subsubsection{CT5}

\begin{figure}
    \centering
    \includegraphics{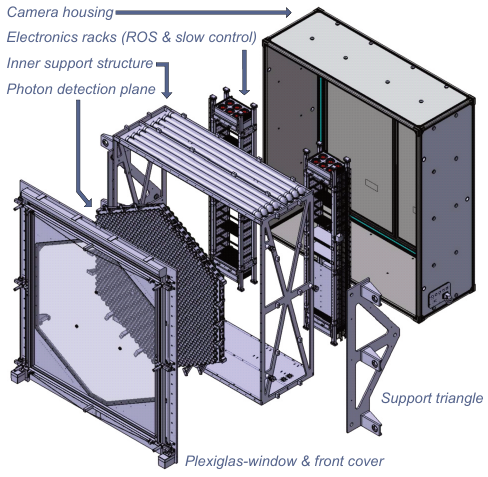}
    \caption{Exploded view of a FlashCam. On the left is the front, pointing towards the telescope reflector. The mechanical building elements are denoted accordingly. (Image adapted from \cite{fc-report-2015})}
    \label{fig:fc-exploded}
\end{figure}

With the currently installed FlashCam, CT5 got a whole new camera in 2019.
The FlashCam concept was introduced in 2008 \cite{FC2008}, and further promoted for CTA e.g. at the 32nd International Cosmic Ray Conference \cite{FC2011}.
The FlashCam concept is highly modular, consisting of separate photon-detection (PDP) and readout electronics (ROS) units, as well as the server based camera data acquisition system (camera DAQ).
PDP and ROS are organized in many, rather small \textit{modules} that can be easily accessed for maintenance.
This paragraph will give an overview of the key elements of a FlashCam as it is installed in CT5.
Its mechanical structure is shown in \autoref{fig:fc-exploded}.\par
The PDP consists of 1758 pixels, arranged in a hexagon.
The pixels are formed of PMTs equipped with Winston cones, each having a field of view (FoV) of \SI{0.08}{\deg}.
The main organization unit of the PDP are its 147 PDP-modules.
Each module holds twelve pixels and support electronics, providing slow-control, HV for the PMTs and a pre-amplifier for each pixel.
Front cover of the PDP is a Plexiglass window that protects the pixels from mechanical damage.
The cover cuts off UV radiation of wavelengths well below \SI{300}{\nano\m}.
It also seals the camera body, so that the system is water- and dust-proof even when the camera-lid is opened.\par
Once a photo-electron (or multiple) induces a signal in one of the PMTs, it gets magnified by a factor of $\sim10$ from the pre-amplifier.
At about \SI{250}{\pe} the pre-amplifier changes its behavior from linear amplification to a quasi-logarithmic one.
In the latter regime, the pulse is clipped and the number of \si{\pe} cannot be determined solely from the signal amplitude anymore, but is determined from the duration of the signal.
This amplification scheme permits a dynamic range from $<\SI{1}{\pe}$ up to \SI{3000}{\pe} with one readout channel per pixel only.\par
The so generated (still analogue) signals are routed to racks at the the rear side inside the camera with commercial shielded, twisted pair cat 6 Ethernet cables.
There the ROS is located.
The ROS consists of about a hundred modules, all based on a common motherboard-design, equipped with different piggybacks, depending on the purpose of the respective module.
Each motherboard has a powerful FPGA (Xilinx Spartan-6) mounted with different code running, depending on the type of piggyback the board carries.
Most modules are used to digitize and process the signals from the PDP.
Such a module then consists of a motherboard and usually two piggybacks equipped with commercial 12-bit flash analog-to-digital converters (FADCs) and is called an FADC-module.
One piggyback handles the signals of twelve pixels and digitizes them at a rate of \SI{250}{\mega Sample\per\s}.
The FPGA on the respective motherboards buffers the whole signal for \SI{32}{\micro\s}.
In case of a trigger, a fixed time window ($\sim\SI{100}{\nano\s}$) is read out to the camera DAQ for further processing.
The continuous signal-digitization allows dead-time free data taking up to \SI{30}{\kilo\Hz} trigger rate.\par
To form the trigger decision, 588 overlapping patches of nine pixels each are defined.
If the sample-wise sum in one of these patches exceeds a given threshold, camera readout is triggered.
To avoid triggering on large single-pixel signals within a patch, the signal of each pixel used for the trigger decision is clipped at a pre-defined, maximum value. 
The trigger decision is fully digital.\par
To enable the formation of the 9-pixel-patches, trigger-modules are used which collect the information of seven FADC-modules each.
They are all connected with each other via a master module that distributes trigger signals over the whole camera.
Additionally, the master module receives and distributes array trigger signals from and to the camera server as well as the precision clock signal from the array infrastructure via WR.\par
When a trigger occurs, the FADC-signals are read out to the camera DAQ via four \SI{10}{\giga Bit\per\s} Ethernet fibers, using a custom developed, high performance Ethernet protocol.
Two further GBit Ethernet fibers are used for slow control and monitoring.
Further, the camera DAQ is responsible for signal processing, i.e.\ to determine the number of \si{\pe} from the signal, either from its amplitude or its duration, and a measure of its arrival time.
The result is sent to the array DAQ to be written on disk.\par
The camera body has dimensions of $3\times3\times1\,\text{m}^3$. 
Besides the above-mentioned components, it hosts a power and safety cabinet and a climate-control system.
The main component of the climate-control system is an active water cooling system which guarantees sufficiently stable temperatures during observations.
A thermocirculator is installed outside the camera in the CT5 telescope structure.
During the first year of observations, the standard deviation of a single pixel's temperature during an observation run\footnote{Operations in \hess\ are organized in \textit{runs}. An observation run typically lasts \SI{28}{\min}. There are other types of runs as well, such as calibration runs with different duration depending on the respective purpose.} was $<$\SI{0.1}{\cel}.
Maximum standard deviation over all pixels was about $<$\SI{1.5}{\cel} \cite{icrc2021-fc-performance}.
In addition, the climate-control system creates a continuous, dry-air over-pressure with respect to the surrounding.
This prevents dust from entering the camera body through potentially remaining small leaks in the sealing and remaining moisture from condensing.
The power consumption of camera and cooling system sums up to about \SI{6}{\kilo\W} \cite{fc-report-2015}.\par
An important feature of FlashCam to be mentioned is its easy maintainability.
If any part of the camera breaks, it can be exchanged within one hour and without dismounting the entire camera.
This is achieved by the mentioned, highly modular concept of the camera.
The rather small modules are easy to access and to exchange.
Nevertheless, during the four years of operations so far in CT5, only a very small number of modules had to be exchanged at all.\par
With the installation of a FlashCam, the availability of the CT5 camera increased significantly to about 98\,\% of the observation time.

\subsection{Central Facilities}\label{CentFac}
The central facilities are essential for data taking as well as the internal and external data transport. They are described in the following.

\subsubsection{Central Trigger System}

The central trigger system (CTS) in \hess\ consists of hardware in a central station located in the control building and of interface modules in each of the five cameras, which are connected to the CTS via optical fibres. The task of the CTS is the implementation of the array trigger as well as the dead time measurement and the synchronisation of events. The implementation of the array trigger is described in the following. The reader is referred to \cite{funk2004} for more information about the \hess\ trigger system. \par

There are two array triggers implemented in \hess\ \textendash\  the stereoscopic and the hybrid trigger. The stereoscopic trigger requires at least two of the CT1-4 telescopes to detect the same event in order to read it out. This reduces significantly the amount of background events written to disk which are mostly caused by muons. Additionally, this leads to a reduced system dead time and allows the usage of a reduced trigger threshold on the camera level leading to a reduction in the system energy threshold. With the addition of CT5 to the array, a new trigger had to be implemented as CT5 has a lower energy threshold than CT1-4. This hybrid trigger only requires CT5 to trigger and additionally any of the smaller telescopes is read out in case it contains a (camera-)trigger signal. CT5 records $\approx 65-70\%$ of the events alone. 
Events recorded with only one of the CT1-4 telescopes are discarded immediately.\par 
Independent of a stereoscopic or a hybrid trigger, the trigger signal is sent to the central station of the CTS if one of the telescopes triggers. The CTS checks for a coincident trigger signal in one of the other four telescopes. To accomplish this, all signals arriving within a short time window are accepted as ``coincident'' (coincidence window). Hereby, the delay caused by the telescope-specific and pointing-dependent light propagation is taken into account. In order to choose the size of the coincidence window, a compromise between too many random telescope coincidences (window too wide) and a loss of valid Cherenkov coincidences (window too small) has to be made. In the end, the size of the coincidence window was chosen to be $\SI{80}{ns}$. If one of the CT1-4 telescopes was the initially triggered telescope, a readout signal is only issued by the CTS in case a coincident trigger signal is found in one of the other telescopes. If not, the event is discarded. For the hybrid trigger (i.e., CT5 being the initially triggered telescope), the event of CT5 and any event of CT1-4 within the coincidence window is read out \cite{funk2004, balzer2014}. \par

\subsubsection{Data Acquisition System}
The data acquisition system (DAQ) is responsible for the coordination of the telescope operations at night. It comprises the hardware and software for storing and merging the data from the cameras, handling the communication between the various subsystems as well as the control and monitoring of the telescope array. During the lifetime of \hess, the DAQ underwent two major upgrades \textendash\ one to incorporate CT5 $\sim 2012$ as described in \cite{balzer2014} and one in $2019$ as described in \cite{zhu2022} in order to ensure safe and reliable operation for the remaining lifetime of \hess\ Here, only a brief overview over the main parts of the DAQ concerning the camera data handling, the DAQ computing hardware and the array control and monitoring is given. For more information, the reader is referred to \cite{balzer2014} and \cite{zhu2022} upon which the following section is based. \par 
The DAQ hardware underwent a full upgrade in 2019 as some of the components had or would have soon reached their end of life status. Therefore, the main goal of the DAQ upgrade was to make the hardware and software future-proof as well as to reduce the long-term maintenance efforts. Another goal was the extension of the storage servers in order to cope with the higher storage needs after the integration of new data taking modes and new/upgraded cameras. \par 
The computing cluster comprises a total of seven computing nodes and three storage servers. The computing nodes have several tasks. At night during observations, six of the seven nodes receive the data from the cameras as well as the trigger information from the CTS. In order to not introduce an additional dead time due to the DAQ, the computing nodes are used in a round-robin load balancing scheme, which is labelled \textit{Node switching} in \hess\ This allows the DAQ to receive and process the data at an higher rate than it is taken. For this \textit{Node switching}, the data of each of the cameras is sent for a short amount of time ($\approx \SI{4}{s}$) to one node during a run. After that time span, the central trigger switches the receiving node and the data is sent to the next computing node. These data chunks are then transformed into the \hess\ raw data format, which is based on the ROOT Data Analysis Framework. Furthermore, while the receiving nodes process the data, they run a preliminary real-time analysis in order to provide feedback to the shift crew. During daytime, a medium-sensitivity data analysis is run on six of the seven nodes (during night on the seventh), while the seventh node serves as login node. After a  run has finished, the data is sent to the three storage servers. \par 

Each of the three storage servers consists of twelve $\SI{8}{TB}$ disks with one of the disks being a spare. The remaining eleven disks of each storage server are combined in a RAID6 configuration resulting in a total of $\SI{188}{TB}$ of disk space available for camera data (after subtracting space for user directories, database storage, ...). After the data of a run is saved, it is transferred to computing clusters in Europe (Lyon, France and Heidelberg, Germany). In case the internet connection to Europe is down, the storage servers provide enough buffer for approximately three months of operations ($40-\SI{50}{TB}$ of data per month). The seven computing nodes and the three storage servers are located in an air-conditioned room in the central control building. Furthermore, the camera servers for the five telescope cameras are placed there. This room is internally known as the ``computer farm''.  \par 

The array control and monitoring is performed from the central control room on the \hess\ site by the shift crew. A total of eight machines are available in the control room, which includes three spares and one machine dedicated to communication. On a total of twelve screens all the relevant information for a successful operation of the telescope array is provided: weather information (e.g. temperature, humidity, cloud cover), camera monitoring parameters (e.g. temperature, high voltage, broken pixels), telescope motion and pointing direction, and real time analysis results (e.g. sky maps). All of this is provided by the DAQ Graphical User Interface (GUI). 
Beside of displaying the aforementioned information, the DAQ GUI enables the shift crew to control the telescopes and schedule the observations of each night. 
Usually, runs are scheduled automatically from a data base. 
This data base is fed with the results of the evaluation of collaboration internal proposals in cycles of one year. 
However, manual rescheduling might be necessary, e.g. to observe targets of opportunity such as transient sources or to perform calibration runs.
Last but not least, the DAQ GUI is the first and most important tool for troubleshooting in case of problems.\par

\subsubsection{Internal and External Network Connection}
The internal network ensures the transfer of data from the cameras to the computing nodes and the storage servers. Additionally, it routes the communication between the various subsystems.
On the other side, the external network connection to the remote \hess\ site is needed for the data transport to the European computing clusters. Furthermore, it provides secured remote access to the various subsystems to the subsystem experts, which are mainly located at institutes in Europe. The currently implemented network configuration is described according to \cite{zhu2022}.\par
The internal network consists of two $\SI{10}{Gb/s}$ and three $\SI{1}{Gb/s}$ switches from which one of the three serves as a spare. The former are connecting all the devices which are involved in the transfer of the data from the cameras to the storage servers. They provide a high throughput to maximize the capabilities of new data taking modes of the cameras. Furthermore, the complete internal network including the DAQ machines, all the subsystems, auxiliary devices and external non-\hess\ devices are separated into various virtual networks (VLANs) in order to provide greater security and stability.\par 
Remote access to the DAQ cluster and the various subsystems of the experiment is of high importance in order to provide the subsystem experts the possibility to access their respective subsystems and perform troubleshooting in case of problems. Starting from 2018, the external network has the responsibility of transferring the data via a, currently, $\SI{100}{Mbps}$ plan to Europe. Before 2018 it was saved on magnetic tapes and then shipped to Europe, as a fast and reliable Internet connection was too expensive. The external access to the DAQ cluster is restricted by a Juniper SRX300 firewall. 

\subsubsection{Power Connection}
The power to the DAQ computing cluster, the telescopes and all electronics on-site is provided by the public power grid. In order to handle possible power interruptions, two diesel generators are used as a backup solution. Furthermore, uninterruptible power supplies (UPSs) are installed to bridge the time of a few seconds between a power loss and the startup of the diesel generators. The UPSs ensure a continuous power supply to the DAQ cluster and the entire array long enough to be able to shutdown and park the telescopes in the case that the diesel generators fail \cite{balzer2014}.

\subsection{Auxiliary Facilities}

There are several auxiliary facilities on the \hess\ site. Some serve to assess observing conditions (weather). Some others are astrophysical-related experiments not directly associated to \hess. This section briefly describes two of the experiments. Other instruments such as a Lidar are just mentioned here.\par 
\subsubsection{ATOM and All-Sky Camera}
The Automatic Telescope for Optical Monitoring (ATOM) is a fully automatic $\SI{75}{cm}$ optical telescope, which is part of the \hess\ experiment and is located on the site. It monitors sources of very high energy $\gamma$-rays, which are visible in the optical waveband, e.g. Active Galactic Nuclei (AGNs). Therefore, it can not only provide multi-wavelength support to \hess\ but also provides alerts in case a source shows higher flux than usual. As the operation of ATOM is fully automatic, an all-sky camera was installed to monitor the cloud cover in order to detect rain and to shutdown ATOM on-time, i.e. before the rain starts. The cloud cover information of this all-sky camera is provided to the \hess\ shift crew and displayed in the control room giving them valuable assistance in assessing the weather situation and observation conditions \cite{atom2015}. \par 
\subsubsection{AERONET}
An instrument assessing the atmospheric aerosols is installed on the \hess\ site since 2016. This instrument is part of the Aerosol Robotic Network (AERONET), which is a project consisting of ground-based photometers distributed all around the world. It measures the aerosol optical depth (AOD) between $\SI{340}{nm}$ and $\SI{1640}{nm}$ by tracking the Sun during the day and the Moon at nights with a moon illumination $>50\%$. The AOD data provided by AERONET can serve as valuable input to assess the atmospheric conditions during \hess\ observations \cite{aeronet1998, atm-corr-scheme}. \par

\subsection{Data Analysis}

\subsubsection{Introduction}
In the following, we give some brief introduction how camera data (time-stamped number of digital counts per pixel) is transformed to science products (spectra, sky maps, light-curves, etc.).\par

For broader comprehensibility, the various steps of data analysis will refer to the data levels (DLs) as explained by \cite{data-formats}.
These data levels represent a common workflow of data reduction for all IACTs in operation, with differences in the detailed implementation.\par
The above mentioned camera output \textendash\ the raw data \textendash\ is referred to as DL0.
In a first step, the digital counts (output of the ADCs) are converted into units of pe. 
Pixels not illuminated by a Cherenkov shower are removed from the \textit{images}\footnote{The representation of an event by number of \si{\pe} in each pixel of one camera is referred to as \textit{image}.} (zero-suppression). 
The results are called  data level 1 (\textit{calibrated data}).
Event building, i.e., the parametrisation of these images with a few geometrical quantities leads to DL2.
Based on these geometrical quantities, DL2 data allows an estimation of how likely the recorded Cherenkov shower (event) was induced by a $\gamma$-ray rather than by a hadron to be computed (estimation of \textit{gammaness}).
In the next step, events that are supposed to be induced by a hadron are rejected. 
The remaining events gathered in a list and supplemented with the functions representing the response of the system (\textit{instrument response functions}, IRFs) are referred to as DL3.
Finally, a statistical analysis leads to the science products of DL4.
In the statistical analysis, the events from the event-list are matched with their respective energy, according to the IRFs.
Additionally, a model for background estimation must be chosen.
Note that some open source tools such as Gammapy \cite{gammapy} split DL4 into multiple data levels to differentiate between various science products.
However, this section focuses on the low-level analysis, i.e., DL0 to DL3.
Thus, the presented steps provide the reader with a good basis for comparison with other IACT facilities.\par

\subsubsection{Data Transfer}
Before the data can be analysed in a publication-ready way, it has to be transferred to Europe, where final data calibration, processing and reduction takes place.
Until 2018, this transfer was conducted by manually bringing the magnetic tapes on which the data was stored to Lyon.
Since then, connection on-site is strong and stable enough to transfer data via the internet.
Given that on average about 1\,TB of data must be transmitted every day \cite{zhu2022} and that the \hess-site is located about \SI{100}{\kilo\m} away from the next city, this is non-trivial.
Via two mobile network towers, data is transferred from site to Windhoek and from there via standard internet transfer to Lyon and Heidelberg, where the further processing takes place.\par

\subsubsection{Data Calibration in \hess}
Data calibration in \hess\ refers to the entire way from DL0 up to DL2.
Two separate chains exist which permit internal cross-checks of data products already at low levels.
An overview is given in \cite{hess-calib} and \cite{crab-obs-hess}.
For the conversion of digital counts from the camera ADCs to \si{\pe}, calibration parameters are used that have either been obtained in the lab before camera installation and are monitored during operations or that are obtained directly on-site.
To proceed to what is called DL1 in \cite{data-formats}, image-cleaning must be applied.
This ensures that only pixels containing Cherenkov light remain in an image.
A standard algorithm (named tail-cut) is that in order to survive the cleaning a pixel must have recorded at least \SI{4}{\pe} and have a neighbouring pixel that has recorded at least \SI{7}{pe} (and vice versa).
For each of these cleaned images, the Hillas parameters \cite{hillas-param} are calculated.
Further, in this step the direction of the shower (i.e., the projection into the sky) and the distance to the shower core (distance from the telescope to the projected impact point
on the ground) are calculated.
The results of this procedure are stored in Data Summary Tables (DSTs), forming a common basis for the more diverse further analysis and corresponding to DL2 in \cite{data-formats}.

\subsubsection{Towards DL3: $\gamma$-Hadron Separation and IRFs}
The probably most critical step in the entire analysis is to separate whether an event was induced by a photon or by a hadron.
In \hess, multiple algorithms are currently in use to move from DL2 to DL3\footnote{Partially, these algorithms also include event reconstruction and are therefore applied at the DL1 stage already.}. 
Beside the standard variant based on the Hillas parameters and described in \cite{crab-obs-hess}, there is the Image Pixel-wise fit for Atmospheric Cherenkov Telescopes (ImPACT) \cite{impact}, a variant of the Hillas analysis using a tree classification method \cite{hillas-tmva}, and a semi-analytical approach named Model++ \cite{m2plus}.
Depending on the type of source, quality and amount of observation time, one additionally has to choose background-rejection cuts which reject individual events based on their reconstructed properties.
Cuts differ by their power of background rejection, normally at the expense of the effectiveness to retain $\gamma$-rays.
Typical \hess\ cut sets are "hard", "standard" or "loose".
Harder cuts reject more background, but \textendash\ given that $\gamma$-hadron separation gets better at higher energies \textendash\ also yield a higher energy threshold.
Loose cuts are therefore often the best choice for sources with soft spectra, like e.g. extragalactic sources suffering absorption from extragalactic background light at higher energies.
Once the algorithm and the cuts are chosen and applied to the data, one receives the DL3 event lists.\par
To finally reconstruct the events, the DL3 data needs to be matched with corresponding IRFs.
They define at which position $\Vec{p}$ and with which energy $E$ a particle will be detected, when it was emitted at its actual position $\Vec{p}_\mathrm{true}$ in the sky and with its actual energy $E_\mathrm{true}$.
Additionally, they give the effective area of the detector as a function of zenith angle and $E_\mathrm{true}$.
IRFs are produced in simulations, combining Monte-Carlo (MC) simulations of the particle showers with detailed simulations of the detector response.
The advantage of such simulations is obvious: while in-field, we do not know the true energy and position of a primary $\gamma$-ray, we have this information in MCs.
Going into details of MC simulations in VHE $\gamma$-ray astronomy is beyond the scope of this chapter.
We refer to chapter \cite{gamma-simus} in this book.\par
While the telescope parameters do not change a lot within short timescales, this is not the case for the main detector volume of an IACT, the atmosphere.
At the \hess\ site, it can change its optical density by an order of magnitude within hours (especially in the biomass-burning season of south-western Africa, i.e., from August to October) and in consequence de- or increase the relative amount of Cherenkov photons reaching the telescopes.
As the simulations leading to the IRFs in use are made under the assumption of one specific model-atmosphere, this effect has to be taken into account.\par
A good and simple parameter of monitoring the atmospheric quality is the (zenith-angle corrected) raw trigger rate, which is dominated by hadrons and therefore constant over the entire sky.
Observations obtained under bad atmospheric conditions often are discarded.
However, given the possibility to derive the atmospheric optical density for each observation run, a correction of these effects is also possible. 
Such a correction allows to use the otherwise discarded data.
Possible analysis approaches are being discussed, e.g. in \cite{atm-corr-scheme}.

\subsubsection{Background Estimation}

With the DL3 event lists and therefore after the $\gamma$-hadron separation, one treats all events remaining in the dataset as $\gamma$-ray candidates.
The direction of the events is reconstructed in the formation of DL2 data, and with the IRFs produced in MC simulations also the energy of the events can be reconstructed.
However, there is still a background present in every observation.
Depending on the strength of the cut level, more or less $\gamma$-like hadrons still remain in the event-list, and diffuse $\gamma$-rays as well as electrons also remain.
To estimate and eventually remove this background, multiple methods exist.
Which method is appropriate for usage depends on the goal of the analysis (sky-map or spectrum) as well as on the source (point-like or extended).
A good and extensive summary can be found in \cite{bkg-modelling}, the key concepts will be briefly mentioned here.\par
The basic idea is to sum up all counts in a test region (on-region) $N_\mathrm{on}$ and subtract the counts from a reference region without a source of known VHE $\gamma$-radiation (off-region) $N_\mathrm{off}$.
A normalization factor $\alpha$ accounts for different sizes of the on- and off-region.
$\alpha$ may depend on the energy, but this shall be ignored for the moment.
This leads to the number of excess counts: $N_\mathrm{excess} = N_\mathrm{on} - \alpha N_\mathrm{off}$.
A complete statistical description of the significance of these excess counts is given in \cite{li-ma}.\par

Defining an off-region is non-trivial, as the \textit{acceptance}\footnote{The (photon-)acceptance states how likely it is that a photon coming into the telescope that would -- following ray-tracing simulations -- be detected at a certain spot of the camera actually is detected and triggers. Due to multiple reasons (mirror focusing to the center of the camera, images close to camera edges not fully contained, etc.), the acceptance is lower at the edges of the FoV.} to $\gamma$-rays as well as to $\gamma$-ray like background events depends on the position in the FoV\footnote{Here, the FoV is the sky field in which $\gamma$-rays can be reconstructed. It differs from the purely geometrical FoV of the telescopes.} of the instrument.\par
A \textit{ring-background} model is traditionally used to create sky-maps. 
It defines as off-region a ring (typically with inner radius $0.5^\circ$ and thickness $0.2^\circ$) around the test region and does principially not rely on particular acceptance geometries.
However, the strongest acceptance variation is radially symmetric (vignetting) and the corresponding background acceptance models are applied also in ring background analyses.
For spectra, the more approriate background model is the \textit{reflected-regions} model.
It works best for \textit{wobble observations}.
In a wobble observation, the center of the FoV is not pointed directly on the source, but with a slight offset.
The offset typically alternates between not more than $\pm1^\circ$ in Declination and Right Ascension.
This is done in order to correct for the different photon-acceptances within the FoV without relying on a background acceptance model.
Off-regions are created by projecting the on-region on a circle whose center is the center of the FoV and the radius is the distance to the center of the on-region.
For extended sources, the \textit{FoV background} model is a good choice.
It fully accounts for the different acceptances of the system in its FoV by normalizing the radial acceptance model of the camera to the entire FoV of the observation.
Caveat of this method is its sensitivity to deviations of the actual system acceptance compared to the model.\par
To not overestimate the background, one must make sure that known sources of $\gamma$-ray emission do not contaminate the region(s) where the background is estimated from.
For the FoV background method, this of course means the entire FoV.
Typically, such known sources get \textit{masked} with simple geometrical shapes, like a circle, a square or an ellipse.
Regions on the sky map covered by this shape are then excluded from the background estimation.

\subsubsection{High-Level Analysis}
Having completed the data reduction, the last remaining step is to obtain the final science products.
This step is referred to as high-level analysis, and includes retrieving spectra, lightcurves and sky maps, for example.
As for any astronomical analysis, the type of object being analysed determines which science products are of interest.
For every analysis except of maps, it is important to define a region of interest, determining from which parts of the FoV the spectrum, lightcurve, etc. shall be computed.\par
While the entire analysis of data including this part was classically performed in closed source tools, available only to members of the collaboration, open source tools became more and more popular in recent years.
In \hess, they can be used for data reduction and analysis from DL3 on. 
Most common nowadays is the Python module gammapy \cite{gammapy}, which is also foreseen as official analysis tool for the upcoming CTA Observatory.
After all, both approaches provide for most cases similar functionality.

\section{Scientific Highlights Achieved with H.E.S.S.}

H.E.S.S.\ has played a leading role in TeV $\gamma$-ray astronomy since it started full operations in 2004. Given the location in Namibia and the large field of view of the H.E.S.S. phase I system, a survey of the Galactic plane around the Galactic center was an obvious project to conduct, which was tremendously successful by massively increasing the known population of TeV $\gamma$-ray sources and by studying the surroundings of the Galactic center. Pointed observations comprised newly discovered sources from the survey as well as well-known Galactic object classes, such as supernova remnants (SNRs), pulsar wind nebulae (PWNe), and stellar binary systems, specifically those hosting a compact object. Also, sensitive observations of the Magellanic clouds became possible with the location of the instrument in the Southern hemisphere. Nevertheless, H.E.S.S. observations have also covered -- with a similar share of observation time -- object types that are equally accessible from the Northern hemisphere, such as dwarf spheroidals, starburst galaxies, active galactic nuclei (AGN) and gamma-ray bursts (GRBs). 

Since many years, the observation program of H.E.S.S.\ has been planned on an annual basis, based on observation proposals which are prepared by all members of the collaboration, and evaluated by committees composed of selected collaboration members. Nevertheless, the system has been designed to be flexible enough to react quickly to new developments in case of need. The observation program comprises unconstrained observations, coordinated monitoring observations fixed in time, as well as target-of-opportunity observations triggered by external information. The H.E.S.S.\ transients follow-up system has evolved substantially over the operational years, and its latest stage is described in \cite{transient-followup-2022}.

In the following, selected highlights of the results achieved with H.E.S.S.\ up to the time of writing of this article are reviewed.

\subsection{Galactic Science}

One of the core goals of VHE $\gamma$-ray observations is to identify celestial objects that contribute substantially to cosmic ray (CR) acceleration. The sensitivity range of telescope systems such as H.E.S.S.\ of $\sim 0.1$ to several 10s of TeV corresponds to emitting particle energies that are a factor of ten higher, i.e.\ $\sim 1$ to several 100s of TeV. Therefore, observations probe a very relevant energy range of the Galactic CR particle population that is accelerated in the observed objects. At the low-energy end, observations connect to particle accelerators in the GeV range where the bulk of Galactic CR energy resides. At the high-energy end, observations probe the accelerators that reach PeV energies,  i.e.\ energies that at least some Galactic CR accelerators should reach to explain the CR spectrum up to its ``knee'' feature at a few times $10^{15}\,$\si{eV}. 

{\bf Galactic Survey:}  The H.E.S.S. I legacy Galactic plane survey (HGPS, \cite{hgps-2018}) has provided for the first time a substantial view on the top potential source classes that are relevant for TeV Galactic CR acceleration: eight supernova remnants (SNRs), 12 firmly identified pulsar wind nebulae (PWNe), eight composite systems (i.e. the identified counterpart is a composite (SNR and PWN) system, one of which (or both) are likely responsible for the TeV emission), and three firmly identified binary systems. Eleven TeV sources have no known counterpart in other wavelengths, and the largest group of sources (36) have associated counterparts but a firm identification was not possible at the time.
 
{\bf Identification:} There are several identification problems with TeV sources. First of all, a firm identification with an astrophysical object known from other wavebands is in many cases very challenging; the fact that many of the new TeV sources were unidentified has triggered a lot of effort in the community to observe these sources with lower-frequency instruments, e.g. using X-ray satellites (e.g. \cite{1834-paper}), starting from the first results published in 2004 \cite{gp-survey-2005, unidentified-2008}. Second, the nature of the particle population seen in TeV $\gamma$-ray observations (either hadrons or leptons) cannot be derived from the TeV spectra alone. While the leptonic interpretation is considered straight forward e.g. in PWN systems, in SNRs or other relevant object classes the identification of hadrons as the emitting particle population is essential for the relevance to their contribution to the CR population, which is dominated by hadronic particles. For H.E.S.S.\ survey sources without known counterparts, the term ``dark sources'' or ``dark accelerators'' was often used, implying that the particles are hadrons which do not emit -- because of the kinematic threshold for $\pi^0$-production -- in frequency bands below GeV energies. Old SNRs for which sensitivities to detect them in lower frequency bands are not sufficient have been speculated to be good counterpart candidates, though proof was elusive.

{\bf Known SNRs:} Young and middle-aged SNRs have been prime targets for H.E.S.S.\ observations since the very beginning. H.E.S.S.\ has provided the first ever resolved $\gamma$-ray image of a SNR, directly proving TeV particle acceleration in the shock of a SNR \cite{rx-j1713-detection}. The corresponding object \textendash\ RX\,J1713.7-3946 \textendash\ has been examined both with further H.E.S.S.\ studies \cite{rx-j1713-2018} and by the community in many follow-up papers, illustrating the whole bandwidth of both opportunities and problems in studying $\gamma$-ray emission from SNRs. The nature of the emitting particle population remains highly debated. H.E.S.S. has provided TeV images and spectra of other SNR where shell emission is seen (e.g.\ Vela Jr. \cite{velajunior-2018}, the remnant of SN\,1006 \cite{sn1006-2010}, RCW\,86 \cite{rcw86-2018}); for all, a hadronic interpretation is possible but not exclusive. Young SNRs (few hundred years old or even younger) are also prime candidates for studies, given that  the onset of acceleration to the highest possible energies in the SNR shock might be seen. While TeV emission from Kepler's SNR has meanwhile been seen \cite{kepler-snr-2022}, TeV photons from other SNRs such as Puppis A or G1.9+0.3 \cite{puppis-a-2015, g1.9-2014} remain undetected, constraining corresponding models. The search was even extended to very young objects, namely core-collapse SNe within a year after explosion \cite{young-cc-sne-2}, since models have predicted particle acceleration even to PeV energies in these objects. 

{\bf New SNRs:} In the HGPS, four sources have been identified which show a clear shell-type morphology: HESS\,J1731-347 \cite{J1731-347}, HESS\,J1534-571, HESS\,J1614-465, and HESS\,J1912+101 \cite{J1534-etal}. For the first two, because of the identification with the corresponding shell-type counterpart in radio continuum emission, an SNR classification has emerged, the latter two remain SNR candidates for the time being. An interpretation as ``dark'' SNRs for the latter two is suggestive, but the identification of counterparts of the first two (in case of HESS\,J1731-347 even a prominent non-thermal X-ray shell) demonstrated that sensitivity or confusion issues might be a relevant factor for finding counterparts for these large (0.5\degree-scale) sources. 

{\bf SNRs \textendash\ Interaction with molecular gas, particle escape:} Besides the lack of an X-ray synchrotron counterpart, the connection (or -- if statistically possible -- spatial correlation) of TeV emission with molecular gas is considered a good way to identify hadronic emission (e.g.\ G349.7-0.2, \cite{G349.7}). Moreover, in many evolved SNRs (where gas interaction might play a relevant role in the morphology), also the spectral connection to GeV spectra (derived with Fermi-LAT) has played a significant role in the hadronic identification efforts. A prominent example is W49B \cite{W49b}, where the GeV spectrum shows a cutoff at low energies which is often considered as a $\pi^0$-induced spectral feature; the smooth spectral continuation towards TeV energies suggests that the TeV emission is also hadronic. Nevertheless, W49B -- similar to other sources with similar features -- displays a soft GeV-TeV spectrum, which probably indicates that substantial escape of the high-energy end of the particle population has already occurred. Such escape during the early phases of SNR evolution (and not only at the end of the Sedov stage) is meanwhile naturally expected in many SNR particle acceleration and evolution models. A key signature to probe this escape is the search of TeV emission from Molecular Clouds in the surroundings of SNRs. H.E.S.S.\ has pioneered such studies using data from W28 \cite{W28-2008}, showing that constraints on the particle diffusion coefficient in the SNR surroundings are implied. On different scales, also RX\,J1713.7-3946 \cite{rx-j1713-2018} or HESS\,J1731-347 \cite{J1731-modeling} are candidates for such investigations.

{\bf Pevatrons:} Objects in which particles are accelerated to PeV energies have been dubbed ``PeVatrons'' \cite{felix-pevatron-2005}. Hopes have been high that TeV $\gamma$-ray spectra extending up to and beyond \SI{100}{\tera\eV} would indicate not only PeV particles in the objects but also their identification as hadrons. However, the detection of $>\SI{100}{\tera\eV}$ emission from the Crab nebula \cite{lhaaso-pev-crab} and interpretation as leptonic Inverse Compton emission has shown that a hadronic interpretation is not straight forward in these cases. Nevertheless, sources with hard, unbroken power-law spectra well beyond \SI{10}{\tera\eV} remain of prime interest as PeVatron candidates. So far, no detected SNR spectrum fulfills this criterion. Several PeVatron candidates have been discovered with H.E.S.S. and have been investigated in detail, e.g.\ HESS\,J1702-420, HESS\,J1826-130, or HESS\,J1641-463 \cite{J1641-pev-cand}, but their nature -- and the confirmation of PeV acceleration -- remain elusive.

{\bf Galactic center:} The Galactic center region is complex for TeV observations, because of a relatively high density of potential TeV emitters. H.E.S.S. observations have revealed for the first time the degree-scale emission from the Galactic ridge, superimposed by two point sources \cite{gal-ridge-2006}. HESS\,J1745-290 is -- taking the diffuse emission separately into account -- a point source that is positionally consistent with the central black hole and its radio counterpart Sgr A* (the other source is the TeV counterpart to the composite SNR G0.9+0.1, likely driven by PWN emission from the object). The nature of the TeV emission from the galactic center is however still unknown. Sgr A* is the most plausible counterpart but conclusive evidence e.g.\ from correlated variability is still elusive \cite{sag-Astar-var-2008}. The diffuse TeV emission which stems from the Galactic center ridge -- proven by its excellent spatial correlation with the gas density from the central molecular zone -- was used to demonstrate that a central accelerator close to the galactic center is responsible for relativistic particles that have penetrated the gas. Most notably, it could be shown that the accelerator was a PeVatron in the past \cite{gc-pevatron-2016}, making this central object the only confirmed hadronic accelerator to PeV energies known so far. The impact of this object to the global energy budget of Galactic CRs is however not clear.

{\bf Superbubbles:} Given that SNR observations have not revealed yet that they are (or have been in the past) accelerating particles to PeV energies, alternative source classes have attracted attention in that regard. Superbubbles -- i.e. the collective phenomenon of massive stellar winds and supernova remnants driving shock fronts -- are an attractive scenario, either accelerating particles from low energies or after injection from pre-accelerators such as the member objects. The TeV emission from Westerlund 1, which is reminiscent of such a scenario is noteworthy in that context \cite{westerlund1-2012}, though again the interpretation of the emission as either hadronic or leptonic is uncertain, and a superbubble scenario is not proven. TeV emission from 30\,Dor\,C in the Large Magellanic Cloud \cite{LMC-TeV-2015} seen with H.E.S.S.\ remains for the moment the only confirmed TeV-emitting superbubble.

{\bf Pulsar wind nebulae: } PWNe constitute the largest source class with clear identification as seen with \hess While acceleration of hadronic particles has been discussed in PWNe, the detected $\gamma$-ray emission is predominantly explained by leptonic interactions of particles injected from the pulsar wind and accelerated at the relativistic termination shocks of PWNe. PWNe are therefore a source of CR electrons, and $\gamma$-ray observations serve to study the particle acceleration and transport on different physical scales of the systems. The Crab nebula stands out for several reasons, not only because it is -- despite its high B-field in the emission zone and therefore comparatively low Inverse Compton-emissivity -- the brightest persistent TeV source in the sky, and is used as calibration source \cite{crab-obs-hess, crab-meyer-2010}. 
The discovery of flares in the GeV domain with Fermi-LAT has cast doubt on the suitability of using Crab as a standard candle in terms of constant flux, but so far there has not been any evidence for significant flux variability in the TeV domain \cite{crab-obs-2013-flare}. Advanced analysis methods employing e.g.\ simulations of the instrument's point spread function (psf) tuned to run-based observation conditions permit that sub-psf extensions of sources observed with \hess\ can now reliably be resolved \cite{sub-psf-res}. The huge photon statistics collected from the Crab nebula has permitted to resolve the TeV-emitting source size to $\sigma \simeq 52''$, significantly incompatible with zero \cite{crab-resolved-2019}.\par 

Contrary to the Crab nebula, most TeV-detected PWNe are dominated by emission from comparatively low B-field regions, where the particles have either been advected to or have diffused into, away from the pulsar. This understanding was to a good extent driven by the discovery of a substantial PWN population in our Galaxy with the H.E.S.S. Galactic plane survey \cite{hgps-2018, pwn-population-2018}. Given the large and often irregular angular extensions of the sources (all substantially more extended than the Crab nebula) and the general offsets of the sources from the powering pulsar (likely because of inhomogeneities of the underlying environment), an identification of the new TeV sources as PWNe is often challenging. In some instances, as association based on the proximity to an energetic pulsar or a known, prominent X-ray PWN was possible \cite{msh15-52-2005, kookaburra-2006, velax-2012}. Only in rare cases \cite{J1825-2005, J1303-2012} was an association of an offset nebula possible based on the energy-dependent morphology of the TeV sources themselves. Such extended nebulae permit to study particle transport (e.g.\ diffusion vs. advection) and environmental conditions, such as magnetic field turbulence away from the pulsar, in great detail \cite{J1825-2006, J1825-2019, velax-2019}. On even larger scales (beyond the region where the pulsar-driven wind determines the dynamics of the medium), over-densities of relativistic electrons have revealed themselves through their $\gamma$-ray emission, and these sources are called PWN halos. While the FoV of \hess\ is small compared to the ground-based arrays with which this $\gamma$-ray phenomenon has been discovered, important contributions can still be expected \cite{geminga-icrc2021}. To that end, a detailed comparison between the \hess\ and HAWC surveys in the overlapping region has been conducted to understand where, e.g., apparent differences between measured source extensions might be related to, e.g. regarding energy dependency and different background determination techniques \cite{hess-hawc-2021}. PWN halos have recently attracted more attention because they might contribute to the population of PeVatron sources discovered with LHAASO.\par

Because of strong energy losses in the ISM, the CR electron spectrum measured at Earth probes the local surroundings of CR electron accelerators. The size of the horizon scales inversely with particle energy. H.E.S.S. measurements of the electron spectrum contribute substantially to the data collected by different balloon and satellite instruments \cite{electrons-2008}, specifically towards the highest energies, where the measured spectrum of electron candidates (above a break at $\sim$ 1\,TeV, which is consistent with other data) continues featureless to well above 10\,TeV \cite{hess-highlights-mdn-2019}. The spectrum can be used to constrain expectations from local accelerators such as nearby pulsars (i.e. Geminga-like PWNe), as well as SNRs.

{\bf Pulsars:} The identification of TeV pulsar emission is only possible through detection of its pulsed emission, using the known ephemerides from lower wavebands. After the Crab pulsar, the Vela pulsar is the second neutron star for which this has been possible, using \hess\ \cite{vela-pulsed-2018}. The Vela pulsed spectrum steeply cuts off at $\sim\SI{100}{\giga\eV}$, and is plausibly the high-energy tail of the pulsed emission seen with Fermi-LAT. A search for further TeV-emitting pulsars as well as the investigation of possible higher-energy pulsed components with \hess\ is ongoing. A noteworthy side remark is that another class of neutron stars, namely soft gamma-ray repeaters, is also under study with \hess, and while no emission has been detected yet from SGR\,1935+2154 \cite{sgr1935-2021}, an extended TeV source colocated with SGR\,1806-20 has been discovered with \hess\ \cite{sgr1806-2018}, the nature of which however is not understood yet.

{\bf Globular clusters:} Globular clusters (GC) are potential TeV sources, given the high source density of millisecond pulsars and other objects, that could give rise to interacting winds and associated shocks. With the exception of HESS\,J1747-248 in close proximity to Terzan 5 \cite{terzan5-2011}, no TeV emission from GC has however been seen with H.E.S.S. \cite{glob-clust-search-2013}. The extended and slightly offset morphology of HESS\,J1747-248 with respect to Terzan 5 is however puzzling. The nature of the particle acceleration in this GC that leads to TeV emission has not been clarified yet.

{\bf Binary systems: } Stellar binary systems that emit $\gamma$-rays can be classified in four categories: Novae, Colliding-wind binaries, Microquasars, and Gamma-ray (loud) binaries. The first three classes are physical definitions, while the fourth is a purely phenomenological definition at present time, still.\par

Novae are thermonuclear bursts at white dwarfs, following episodes of accretion from the companion. The fact that such outbursts emit $\gamma$-rays in the GeV band has been known since the Fermi era. TeV emission has meanwhile also been detected from the 2021 outburst of the recurrent nova RS Ophiuchi with H.E.S.S. \cite{rs-oph-2022}, as well as with MAGIC and CTA-LST1. The emission is generally interpreted as being of hadronic nature. The particle acceleration originates from the interaction of explosion ejecta into the companion's wind, and the fast evolution of the environment and corresponding acceleration conditions provides an ideal testbed to study particle acceleration in situ, which can be transferred e.g. to SNR shock acceleration in dense environments. \par
In colliding-wind binaries, on the other hand, the acceleration region is thought to be created by the collision of two stellar winds. A modulation of the $\gamma$-ray emission with the binary orbital phase is expected. For $\eta$ Car, such emission and behaviour has indeed been seen with Fermi-LAT in the GeV band \cite{eta-car-fermi-2010}. The VHE source discovered with H.E.S.S., up to $\sim$ 400\,GeV, from the $\eta$ Car position \cite{eta-car-detection-2020} is consistent with emission from the colliding wind region, a confirmation with the measurement of modulation is however still pending. \par
Microquasars are defined through the existence of relativistic jets, powered by the accretion onto the compact object. The compact object can be a black hole, but a neutron star is possible as well. Emission from several microquasars has been reported in the high-energy band using Fermi-LAT. With \hess, persistent TeV emission from the extended lobes of SS 433 can be seen\footnote{\url{https://indico.icc.ub.edu/event/46/contributions/1302/}}, consistent with the discovery of $\gamma$-ray emission with the HAWC experiment. The higher angular-resolution data from \hess\ promise to constrain the particle acceleration sites (likely electrons) that are able to accelerate to PeV energies (seen with HAWC). Other microquasar systems (GRS\,1915+105, Circinus X-1, and V4641\,Sgr) have been observed but so far no TeV emission has been detected \cite{hess-microquasars-2017}. Also newly detected systems such as MAXI\,J1820+070 during its 2018 outburst have been observed, yielding constraining upper limits \cite{maxi-J1820-2018}.\par

The class of Gamma-ray (loud) binaries (GRLBs) is solely defined by the observation that the photon energy distribution measured from these systems peaks above 1\,MeV. Many of these systems (of which currently of order 10 are known, including a few candidates) have therefore been discovered in the $\gamma$-ray bands with Fermi-LAT or \hess, but several were also known from lower frequency band observations. For three (possibly four) of these systems, the compact object is known to be a pulsar, and the emission is likely associated to the interaction of the pulsar wind with the massive companion star. It is suggestive to assume that all GRLBs share this setting, but black holes as compact objects are currently not excluded for several of these systems. Known orbital periods range from 3.9 days to $\sim$ 50 years, and observations of GRLBs hosting a pulsar hold the promise of not only studying wind-wind interactions and associated processes, but also potentially probing pulsar winds on different (including small) scales, since the orbital parameters largely vary. \hess\ was used to observe (and detect) the prototypical system PSR\,B1259-63 soon after begin of operations to study its $\gamma$-ray emission around all periastron passages ever since \cite{B1259-2005, B1259-2009, B1259-2013, B1259-2020}. The photon statistics acquired from LS\,5039 even permitted to derive the binary period (3.9 days) purely from TeV observations with very high accuracy \cite{LS5039-2006}. While the point-like appearance of the TeV source HESS\,J1832-093 (together with an optical counterpart) was already suggestive of a binary nature of the source \cite{J1832-2014} before the orbital modulation could be identified in X-rays, other binary (candidates) suffer from contamination of superposed extended TeV sources. For 1FGL\,J1018.6-5856 the binary nature of the TeV counterpart is confirmed through its modulation \cite{J1018-2015}, but for 4FGL J1405.1-6119 a clear identification of the TeV emission which is detected with H.E.S.S.\ as stemming from the binary is still elusive, given the lack of modulation and the overlap with an extended TeV source.

\subsection{Extragalactic Science}

A scanning survey of a substantial fraction of the extragalactic sky was not within reach for H.E.S.S., despite its relatively large field of view. Extragalactic targets broadly fall into two main categories: Galaxies whose emission is dominated by the central supermassive black hole (Active Galactic nuclei, AGN) and Galaxies in which stellar-size phenomena -- similar as in our Galaxy -- can be studied. Phenomena on Galaxy cluster scale are also probed, H.E.S.S. observations of the Coma cluster, Abell 496, and Abell 85 have however only yielded upper flux limits \cite{coma-upper-2009, abells-upper-2009}. Transient phenomena have constituted a big part of the extragalactic observation program, driven by AGN but also e.g. by gamma-ray bursts (GRBs) which fall under the stellar-size phenomena, or by multi-messenger triggers such as neutrino or gravitational wave events recorded from the sky.

{\bf Magellanic clouds: } 
The Large Magellanic Cloud (LMC) is so far the only celestial object in which stellar-sized objects outside of our Galaxy can be resolved in TeV $\gamma$-rays. With H.E.S.S., so far one object each of the following categories has been detected in the LMC, representing extreme cases of the respective source class: 30 Dor C (a superbubble) \cite{LMC-TeV-2015}, N 132 D (an SNR) \cite{LMC-TeV-2015, N132D-2021}, N 157 B (a PWN) \cite{LMC-TeV-2015}, and LMC P3 (a $\gamma$-ray binary) \cite{LMC-P3-2018}. So far, no VHE emission from SN 1987a has been detected \cite{LMC-TeV-2015}. No detection of VHE emission from the SMC has been reported.

{\bf Starburst galaxies: } 
The investigation of the ensemble of CRs in our Galaxy through its VHE radiation is very challenging, because of a large contribution of unresolved sources and the limited FoV of VHE instruments \cite{diff-gr-2014}. The detected TeV emission from the nearby starburst galaxy NGC 253, on the other hand, could -- in conjunction with the GeV spectrum from Fermi-LAT -- best be interpreted as the emission of the diffuse relativistic proton population in that galaxy, naturally with a CR density well beyond the Galactic one \cite{NGC253-2018}.

{\bf AGN -- Radio galaxies: } 
VHE emission from two prominent, nearby radio galaxies has been detected with H.E.S.S.: Centaurus A and M87. The emission is usually attributed to particles in the jets, though they have low Doppler factors ($\sim 1$). The VHE emission from Centaurus A is in fact constant and spatially resolved and traces the inner parts of the radio lobes, and is best interpreted as coming from ultrarelativistic electrons in the jets \cite{CentA-2020}. Contrary to that, the emission from M87 is highly variable, exhibiting flares down to days timescales \cite{M87-2006, M87-2012}, and with evidence for emission even at the base of the jet \cite{M87-2009}. A recent highlight was the participation of H.E.S.S. together with many other facilities of the electromagnetic spectrum in a M87 observation campaign in conjunction with Event Horizon Telescope (EHT) measurements, with the aim of characterizing the broadband behavior of the source at the time of the EHT black hole shadow measurements \cite{M87-2021}.

{\bf AGN -- population and EBL: } 
With increasing distance (redshift $z$), blazars quickly start to dominate the extragalactic source population detected with VHE instruments (with M87 already also being classified as a ``misaligned'' blazar). The absorption of VHE $\gamma$-rays through interaction with the extragalactic background light (EBL) limits the horizon to about $z \sim 1$. Actually, VHE observations of blazars are used to constrain the level of extragalactic background light. H.E.S.S. pioneered these studies by showing that the EBL was lower than previously thought, using the measured spectra of 1ES 1101-232, H 2356-309 and 1ES 0229+200 \cite{ebl-blazar-2006, blazar-2007, ebl-blazar-2007}. Measurements of a larger number of blazars with known redshifts in fact permit to directly measure the EBL level (i.e. without EBL model scaling), only assuming that the intrinsic VHE spectra are described by smooth concave shapes \cite{ebl-blazar-2013, ebl-sed-2017}. For blazars with unknown redshift, VHE spectra (in combination with GeV spectra from Fermi-LAT) can also be used to give constraints on the redshift of these sources \cite{blazar-redshift-2020}. 

While a dedicated extragalactic survey has not been performed with H.E.S.S., given the large FoV of CT1-4, H.E.S.S. has during its pointed observations also observed a substantial fraction of the extragalactic sky (6\% until 2013, H.E.S.S.\ extragalactic scan (HEGS)\footnote{\url{https://www.mpi-hd.mpg.de/hfm/HESS/pages/home/som/2019/03/}}). Data can be used for population studies, transient searches, and variability searches, and for some of $\sim$ 200 Fermi-LAT sources serendipitously covered, the respective VHE non-detection can be used to constrain the spectral shape (with the need to take biases because of the non-simultaneity of the observations into account).

{\bf AGN -- blazars: } 
While BL Lacs provide the somewhat cleaner environment to study jet physics, flat-spectrum radio quasars (FSRQs) are particularly interesting (and challenging) because of the increased complexity when interpreting their spectra. Besides EBL absorption, their VHE emission can also be affected by absorption from broad-line region emission. Examples are PKS 0736+017 and 3C 279 \cite{pks0736-2020, 3c279-2019}, whose spectra have been examined also with the aim to put constraints on the location of the emitting regions within the jets. Low-frequency peaked AGN are also meanwhile studied with H.E.S.S. (ApLib and PKS 1749+096, \cite{aplib-2015, ToO-blazars-2017}). \par
The VHE emission of selected (in particular hard-spectrum) sources can be used to probe the density of the surrounding intergalactic magnetic field \cite{igm-blazar-2023}.

{\bf AGN -- variability studies: } 
A lot of information can be deduced from variability studies of blazars, in particular when possible in conjunction with observations in other wavebands (in particular X-rays, but also e.g. in the optical). Variability is potentially a combination of line-of-sight (Doppler boost) effects and intrinsic emission variability. A highlight was the observation of a giant flaring episode from PKS 2155-304 in 2006 \cite{pks2155-2007}, with burst timescales down to $\sim\qty{200}{\s}$. PKS 2155-304 has remained a prime target for H.E.S.S. to be studied since the beginning of the experiment \cite{pks2155-2020, pks2155-2017, pks2155-2014, pks2155-2012, pks2155-2010, pks2155-2009, pks2155-apj-2009, pks2155-2005}. Another prime target in this context it the FSRQ PKS 1510-089 \cite{pks1510-2023, pks1510-2021}, whose variability and the correlation to lower frequency bands shows complex behaviour which might also help to understand a phenomenon called ``orphan flare'' (i.e. a VHE flare unaccompanied by variability in lower frequency band observations). \par
Strong flares of sources (optimally at larger redshifts) are also the basis for constraining Lorentz invariance violation (LIV). The 2015 flare observed from 3C 279 (z = 0.54) was a particularly good showcase in that respect \cite{3c279-2019}.

{\bf Gamma-ray bursts: } 
H.E.S.S. has started a gamma-ray burst (GRB) follow-up program soon after its start of operations. Over the years, an automatic pipeline (ultimately with automatic repointing of the telescopes) has been implemented to follow-up GRBs as soon an possible after the burst alert from the prompt phase.
It took however over a decade of hunting for VHE emission from GRBs, e.g. \cite{grbs-2009, grbs-2014} before efforts paid off. In 2018, both MAGIC and H.E.S.S. detected for the first time VHE emission from (different) GRBs. The H.E.S.S. observations of GRB 180720B \cite{grb-180720b-2019} and of GRB 190829A \cite{grb-190829-2021} have revealed VHE emission actually from deep in the afterglow phase of the respective GRBs. 

{\bf Multi-messenger and transient searches: } 
Besides the extensive observation programs triggered by flaring AGN and by GRBs, H.E.S.S. also follows a variety of other triggers issued by detections in the electromagnetic (EM) spectrum as well as from other messengers, specifically neutrinos and gravitational waves. Triggered by EM-induced alerts, Galactic transients such as microquasars or other X-ray binaries are followed up if they meet predefined trigger criteria. Gravitational wave alerts have been followed up since the start of LIGO/Virgo operations, following trigger alerts which have initially been shared only amongst EM partners. H.E.S.S. has participated in the joint EM follow-up observations of GW 170817, the first gravitational-wave event (a binary neutron star merger) with confirmed electromagnetic counterparts (though not in VHE) \cite{gw170817-2017}. Follow-up observations of binary black hole mergers are discussed in \cite{bbh-followup-2021}. \par
Regarding neutrino follow-up observations with H.E.S.S., the most noteworthy event was IceCube-170922A which has been associated with the blazar TXS 0506+056, where in fact the MAGIC telescopes have revealed a VHE signal \cite{ic170922a-2018}. In general, the interpretation of the results (specifically emission episodes at different times) is challenging because of the different opaquenesses of typical objects to $\gamma$-rays and neutrinos.

\subsection{Dark or Exotic Matter Searches}
Dark matter particles may reveal themselves through self-annihilation or decay and subsequent $\gamma$-ray emission. The annihilation process may lead to line-like features in the spectra. Over the years, different (dedicated) targets or surveys have been investigated. Target selection is such that the expected dark-matter density is high while the emission from astrophysical sources should be low. Investigated sources/fields comprise the central region of the Milky Way \cite{DM-igps-2022, DM-GalCent-2016, DM-GalCent-2015}, the Galactic halo \cite{DM-GalHalo-2018, DM-GalHalo-2016}, dwarf spheroidals or irregulars \cite{DM-DIrr-2021, DM-DSpher-2020, DM-WIMP-2018, DM-DSpher-2014, DM-DSpher-2011}, unidentified Fermi objects \cite{DM-UFO-2021}, globular clusters \cite{DM-GC-2011}, and even from galaxy clusters \cite{DM-Fornax-2012} and intermediate black holes \cite{DM-IMBH-2008}. The derived upper limits constrain the parameter space for possible dark matter masses and annihilation cross-sections.

Other exotic particle searches concern axions (which could be dark matter candidate particles). Such particles may modify the $\gamma$-ray spectra from distant sources. For example, constraints from potential imprints on the spectrum of PKS 2155-304 have been reported in \cite{DM-PKS2155-2013}.

\section{Conclusion}

\hess\ has been and continues to be a key TeV facility, providing excellent scientific results to the astrophysical community and beyond. 
There are strong reasons for a potential continuation of operations beyond the current (2024 at the time of writing) extension phase. For several years to come, \hess\ may be the only sensitive TeV facility with good access to the Southern sky.

\newpage

\printbibliography[heading=bibintoc]

\end{document}